\documentclass[adp,fleqn]{w-art}
\usepackage{times}
\usepackage{w-thm}
\usepackage[]{graphicx}

\chardef\bslash=`\\ 

\newcommand{\vp}{\varphi}
\def\be{\begin{equation}}
\def\ee{\end{equation}}
\def\bea{\begin{eqnarray}}
\def\eea{\end{eqnarray}}
\def\ba{\begin{eqnarray}}
\def\ea{\end{eqnarray}}
\newcommand{\rd}{{\rm d}}

\newcommand{\lsim}{\mbox{\raisebox{-1.ex}{$\stackrel
     {\textstyle<}{\textstyle \sim}$}}}
\def\R{{\cal R}}

\def\xrightarrow#1#2#3#4{\,\lower#1pt\hbox{$\stackrel{\stackrel{\displaystyle #2}%
{\hbox to #3cm{\rightarrowfill}}}{#4}$}\,}

\begin{document}

\DOIsuffix{theDOIsuffix}


\pagespan{1}{}


\title[Cosmologies from higher-order string corrections]
{Cosmologies from higher-order string corrections}

\author[Shinji Tsujikawa]{Shinji Tsujikawa\footnote{Corresponding
     author: e-mail: {\sf shinji@nat.gunma-ct.ac.jp}, }\inst{1}} 
\address[\inst{1}]{Gunma National College of Technology,  
580 Toriba,  Maebashi, Gunma 371 8530, Japan}

\begin{abstract}
    
We study cosmologies based on low-energy effective string theory 
with higher-order string corrections to a tree-level action 
and with a modulus scalar field 
(dilaton or compactification modulus). 
In the presence of such corrections it is possible to construct 
nonsingular cosmological solutions 
in the context of Pre-Big-Bang and Ekpyrotic universes.
We review the construction of 
nonsingular bouncing solutions 
and resulting density perturbations 
in Pre-Big-Bang and Ekpyrotic models. 
We also discuss the effect of higher-order string corrections on
dark energy universe and show several interesting possibilities of
the avoidance of future singularities.

\end{abstract}
\maketitle              

\section{Introduction}

String theory has continuously stimulated its application to 
cosmology in a number of profound ways \cite{Lidsey99,Mariusz,Gasperini02}.
It is actually very important to test the viability of string 
theory by extracting cosmological implications from it.  In particular 
string cosmology has an exciting possibility to resolve the big-bang 
singularity which plagues in General Relativity.

The Pre-Big-Bang (PBB) model \cite{Veneziano91,pbb} 
based on the low energy, tree-level string effective action 
is one of the first attempts to apply string theory to 
cosmology.
In this scenario there exist two disconnected branches, one of which corresponds 
to the dilaton-driven inflationary stage and another of which is the 
Friedmann branch with a decreasing curvature.  Then string corrections to the 
effective action can be important around the high-curvature regime where 
the branch change occurs.  Ekpyrotic/Cyclic cosmologies 
\cite{ekpyr,ekpyr2,cyclic} have a similarity to the PBB scenario in the sense that 
the description in terms of the tree-level effective action breaks down around 
the collision of two branes in a five dimensional bulk.

When the universe evolves toward the strongly coupled, high-curvature regime 
with growing dilaton, it is inevitable to implement higher-order 
string corrections to the tree-level action.  Indeed it was found that two branches can be 
smoothly joined by taking into account the dilatonic higher-order 
string corrections in the context of PBB \cite{Gas96,Bru} and 
Ekpyrotic \cite{TBF02} scenarios.  In the system where 
a (compactification) modulus 
field is dynamically important rather than the dilaton, 
Antoniadis {\it et al.} showed that the big-bang singularity can be avoided 
by including the Gauss-Bonnet (GB) curvature 
invariant coupled to the modulus \cite{Anto93}.

In order to test these string-motivated models from observations, 
it is important to investigate spectra of density perturbations and to compare
them with temperature anisotropies in Cosmic Microwave Background (CMB).
For example, inflationary cosmology generically predicts nearly 
scale-invariant spectra of density perturbations.
This prediction agrees well with the recent observations in 
CMB anisotropies \cite{WMAP}.
While inflationary cosmology is based upon the potential energy of 
a scalar field with a slow-roll evolution, the kinematic energy of 
dilaton or modulus field dominates in PBB and Ekpyrotic/Cyclic 
cosmologies. Hence it is expected that the spectrum of 
density perturbations in the latter case is different from 
the prediction in inflationary cosmology.
We shall address the problem of density perturbations
generated in PBB and Ekpyrotic/Cyclic 
cosmologies by using nonsingular bouncing solutions
obtained by including second-order string corrections.
We note that the effect of string corrections can be 
important in constructing inflation models, 
see Refs.~\cite{inflationworks} for such possibilities.

The effect of such string corrections can be also important 
in the context of dark energy.
{}From recent observations the equation of state 
(EOS) parameter $w$ of dark energy lies
in a narrow strip around $w=-1$ quite likely being below of
this value \cite{SN,obser} (see Refs.~\cite{reviewdark} for 
reviews of dark energy).
The region where the EOS parameter $w$
is less than $-1$ is referred as 
a phantom (ghost) dark energy universe \cite{Caldwell}.
The phantom dominated universe ends up with a
finite-time future singularity called Big Rip 
or Cosmic Doomsday \cite{CKW,Carroll,Singh}.
The Big Rip singularity is characterized by divergent behavior 
of the energy and curvature invariants at Big Rip time.
Hence it is natural to account for higher-order 
curvature corrections in the presence of dark 
energy \cite{NOT,NOS,Topo,CTS,ACD,Neu,Cog,Cal06,Are,NOS2,KM06}.
In fact it is possible to
avoid or moderate the Big Rip singularity
when such corrections are present \cite{NOT,Topo,CTS} 
and showed that it is possible to
avoid or moderate the Big Rip singularity.
We shall review cosmological solutions in second-order
string gravity in the context of dark energy and consider the 
avoidance of future singularities.

In what follows the effect of string corrections will be reviewed
in two separate sections--(i) PBB/Ekpyrotic cosmologies (sec.\,2) and 
(ii) dark energy universe  (sec.\,3).
It is interesting to note that such corrections can play 
important roles for both past and future singularities.

\section{Pre-Big-Bang and Ekpyrotic cosmologies}

The PBB scenario is based upon low-energy, tree-level 
effective string theory using  toroidal compactifications \cite{Veneziano91,pbb}.
The string effective action in four dimensions is given by 
\be
 \label{effactions}
{\cal S}_S \, = \, \int {\rm d}^4x \sqrt{-g} e^{-\phi}
\left[ \frac12 R+\frac12 (\nabla \phi)^2-V_S(\phi)+
{\cal L}_c+{\cal L}_m \right]\,,
\ee
where $\phi$ is a dilaton field that controls the string 
coupling parameter, $g_{s}^2=e^{\phi}$, and 
$\sqrt{-g}$ is the determinant of metric $g_{\mu \nu}$.
We neglect here additional modulus fields corresponding to
the size and shape of the internal space of extra dimensions.
The potential $V_S(\phi)$ for the dilaton 
vanishes in the perturbative string effective action.
The Lagrangian ${\cal L}_c$ corresponds to higher-order
string corrections which we will present later, 
whereas ${\cal L}_m$ is the Lagrangian of additional 
matter fields (e.g., fluids, kinetic components, axion etc.). 
In this section we do not consider the contribution of 
${\cal L}_m$, but in Sec.~\ref{dark} we will account for it 
as a barotropic fluid.
The above action is so-called the ``string frame'' action 
in which the dilaton is coupled to a scalar curvature, $R$.

The dilaton $\phi$ starts out from a weakly coupled regime ($g_s \ll 1$) 
and evolves toward a strongly coupled region ($g_s>1$).
The Hubble parameter grows during this stage.
This ``superinflation'' is driven by a kinematic energy of 
the dilaton field, which is so-called a PBB branch. 
There exists another Friedmann branch 
with a decreasing curvature. 
It is possible to connect the two branches
by accounting for higher-order string corrections ${\cal L}_c$ to
the tree-level action \cite{Gas96,Bru,Foffa99,Cartier99}. 
This is one of the main topics in this review.

In Ekpyrotic \cite{ekpyr,ekpyr2} and Cyclic \cite{cyclic} cosmologies
the universe contracts before the bounce because of the presence of 
a negative potential characterizing an attraction force between 
two parallel branes in an extra-dimensional bulk.
The collision of two parallel branes 
signals the beginning of the hot,
expanding, big bang of standard cosmology.
After the brane collision the universe connects to a standard 
Friedmann branch as in the case of PBB cosmology.
The origin of large-scale structure is supposed to be generated
by quantum fluctuations of a field $\phi$
characterizing the separation of a bulk brane.
It is important to construct nonsingular bouncing cosmological solutions
in order to make concrete prediction of the power spectrum 
generated in Ekpyrotic/Cyclic cosmologies.
This is actually possible by accounting for higher-order string 
corrections as in the PBB case \cite{TBF02}.

The PBB model has a similarity to Ekpyrotic/Cyclic cosmologies
in a sense that the universe exhibits a bounce in 
``Einstein frame''. 
Making a conformal transformation
\be
 \label{conformalframe}
\hat{g}_{\mu\nu}=e^{-\phi}g_{\mu\nu}\,,
\ee
the action in Einstein frame is given by 
\be \label{effactione}
{\cal S}_E \, = \, \int {\rm d}^4x 
\sqrt{-\hat{g}}\left[ \frac12 \hat{R}
- \frac14 (\hat{\nabla} \phi)^2-V_E(\phi)
+\cdots \right]\,,
\ee
where $V_E(\phi) \equiv e^{\phi}V_S(\phi)$.
Introducing a rescaled field $\varphi=\pm \phi/\sqrt{2}$,
the action (\ref{effactione}) yields
\be \label{effactione2}
{\cal S}_E \, = \, \int {\rm d}^4x 
\sqrt{-\hat{g}}\left[ \frac12 \hat{R}
-\frac12 (\hat{\nabla} \varphi)^2-V_E(\phi(\varphi))+
\cdots \right]\,.
\ee
This is the action for an ordinary scalar field $\vp$
with potential $V_{E}$.
Hence it can be used to describe both the
PBB model in Einstein frame, as well as the ekpyrotic 
scenario \cite{Durrer02}.

In the original version of the Ekpyrotic scenario \cite{ekpyr}, the
Einstein frame is used where the coupling to the Ricci curvature is
fixed, and the field $\phi$ describes the separation of a bulk brane
from our four-dimensional orbifold fixed plane.  In the case
of the second version of the Ekpyrotic scenario \cite{ekpyr2} and in
the cyclic scenario \cite{cyclic}, $\phi$ is the modulus field
denoting the size of the orbifold (the separation of the two orbifold
fixed planes).

The Ekpyrotic scenario is described by a negative
exponential potential \cite{ekpyr}
\be \label{einpoten}
V_E = - V_{0} \exp
\left(-\sqrt{\frac{2}{p}}\,\varphi\right)\,,
\ee
with $0 < p \ll 1$. 
The branes are initially widely separated but are approaching each 
other, which means that $\varphi$ begins near $+\infty$ and is
decreasing toward $\varphi = 0$.
In the PBB scenario the dilaton starts to evolve from 
a weakly coupled regime with $\phi$ increasing from $-\infty$.
If we want the potential (\ref{einpoten}) to describe a
modified PBB scenario with a dilaton potential which is important
when $\phi \to 0$ but negligible for $\phi \to - \infty$, we
have to use the relation $\varphi=-\phi/\sqrt{2}$ between the field
$\varphi$ in the ekpyrotic case and the dilaton $\phi$ 
in the PBB case.

In the flat Friedmann-Robertson-Walker (FRW) metric
${\rm d}s^2=-{\rm d}t_{E}^2+a_E^2 {\rm d}x_{E}^2$
in Einstein frame, the background equations with 
${\cal L}_c=0$ are given by 
\begin{eqnarray}
& &3H_{E}^2=\frac12  \dot{\vp}^2+V_E (\vp)\,,\\
& &\ddot{\vp}+3H_E\dot{\vp}+\frac{{\rm d}V_E}
{{\rm d}\vp}=0\,,
\end{eqnarray}
where a dot represents a derivative with respect to $t_E$
and $H_E \equiv \dot{a}_{E}/a_{E}$.
Here the subscript ``{\it E}\,'' denotes the quantities
in the Einstein frame.
The exponential potential (\ref{einpoten}) has the following exact
solution \cite{LM85}
\be 
\label{ekysolution}
a_E \propto |t_E|^p, \quad H_E=\frac{p}{t_E},\quad
V_E=-\frac{p(1-3p)}{t_E^2},\quad
\dot{\varphi}=\frac{\sqrt{2p}}{t_E} \,.
\ee
The solution for $t_E<0$ describes the contracting universe
prior to the collision of branes.
The Ekpyrotic scenario corresponds to a slow contraction
with $0<p \ll 1$.
{}From Eq.~(\ref{ekysolution}) the potential vanishes for 
$p=1/3$, which corresponds to the  PBB scenario.

In string frame the scale factor $a_{S}$ and the
cosmic time $t_{S}$ are related with those in 
Einstein frame via the relation
${\rm d}t_{S}=e^{-\varphi/\sqrt{2}}{\rm d}t_E$ and 
$a_{S}=e^{-\varphi/\sqrt{2}}a_E$.
Then we find \cite{TBF02,Durrer02}
\be
\label{ekysolution2}
 a_S \propto (-t_S)^{-\sqrt{p}},\quad
\phi=-\frac{2\sqrt{p}}{1-\sqrt{p}}
\ln \left[ -(1-\sqrt{p})t_S\right] \,.
\ee
This illustrates a super-inflationary solution with growing dilaton.
Hence the contraction in Einstein frame corresponds to the 
superinflation driven by a kinematic energy of the field $\phi$.
We note that there exists another branch of an accelerated contraction
($a_S \propto (-t_S)^{\sqrt{p}}$) \cite{Gasperini02}, 
but this is out of our interest.

The above solution needs to be regularized around $t_S=0$ (or $t_E=0$) 
in order to connect to the Friedmann branch after the bounce.
In the context of PBB cosmology, it was realized
that higher-order string corrections 
(defined in the string frame) to the action induced by inverse
string tension and coupling constant corrections can yield a
nonsingular background cosmology.
A possible set of corrections include terms of the 
form \cite{Gas96,Bru}
\begin{eqnarray}
 {\cal L}_c = -\frac12 \alpha' \lambda
  \zeta(\phi) \left[ c R_{\rm GB}^2+ d
 (\nabla \phi)^4 \right]\,,
\label{lagalpha}
\end{eqnarray}
where $\alpha'$ is a string expansion parameter, 
$\zeta(\phi)$ is a general function of $\phi$ and $R_{\rm GB}^2
=R^2-4R^{\mu\nu}R_{\mu\nu}+
R^{\alpha\beta\mu\nu}R_{\alpha\beta\mu\nu}$ is
the Gauss-Bonnet (GB) term.
$\lambda$ is an additional parameter which depends on 
the types of string theories: $\lambda=-1/4$, $-1/8$ and 0
correspond to bosonic, heterotic and superstrings, respectively.
If we require that the full action agrees with the three-graviton 
scattering amplitude, the coefficients $c_{i}'s$ are fixed to be
$c=-1$ and $d=1$ with $\zeta(\phi)=-e^{-\phi}$ \cite{Met}.

The corrections ${\cal L}_c$ are the sum of the tree-level
$\alpha'$ corrections and the quantum $n$-loop corrections
($n=1, 2, 3,\cdots$), with the function $\xi(\phi)$ given by
\be
\zeta(\phi)=-\sum_{n=0} C_n e^{(n-1)\phi} \,,
\label{xifunction}
\ee
where $C_n$ ($n \ge 1$) are coefficients of
$n$-loop corrections, with $C_0=1$.
There exist regular cosmological solutions 
in the presence of tree-level and one-loop corrections, 
but this is not realistic in the sense that the Hubble 
rate in Einstein frame continues to increase 
after the bounce (see Fig.~1 of Ref.~\cite{TBF02}).
Nonsingular bouncing solutions that connect to a Friedmann
branch can be obtained by accounting for the corrections up
to two-loop with a negative coefficient ($C_2<0$).

It was shown in Ref.~\cite{TBF02} that 
nonsingular bouncing solutions exist in Einstein frame 
even in the presence of a negative exponential
potential.  When $p \ll 1$ the potential is vanishingly small for 
$\varphi \gg 1$, in which case the dynamics of the system is practically 
the same as that of the zero potential discussed in 
Ref.~\cite{Bru}.
In this case the dilaton starts out from the low-curvature regime $|\phi| 
\gg 1$, which  is followed by the string phase with linearly growing dilaton 
and nearly constant Hubble parameter.
During the string phase one has \cite{Gas96} 
\begin{eqnarray} 
a_S \propto (-\eta_S)^{-1},\quad
\phi=-\frac{\dot{\phi}_f}{H_f} {\rm ln} (-\eta_S)
+{\rm const}\,,
\label{const}
\end{eqnarray}
where $\dot{\phi}_f \simeq 1.40$ and $H_f \simeq 0.62$.
In the Einstein frame this corresponds to a contracting Universe with
\begin{eqnarray} 
a_E \propto (-\eta_E)^{\dot{\phi}_f/(2H_f)-1}\,.
\label{scaleein}
\end{eqnarray}

\begin{figure}[htb]
 \begin{center}
\includegraphics[width=9cm, height=15cm]{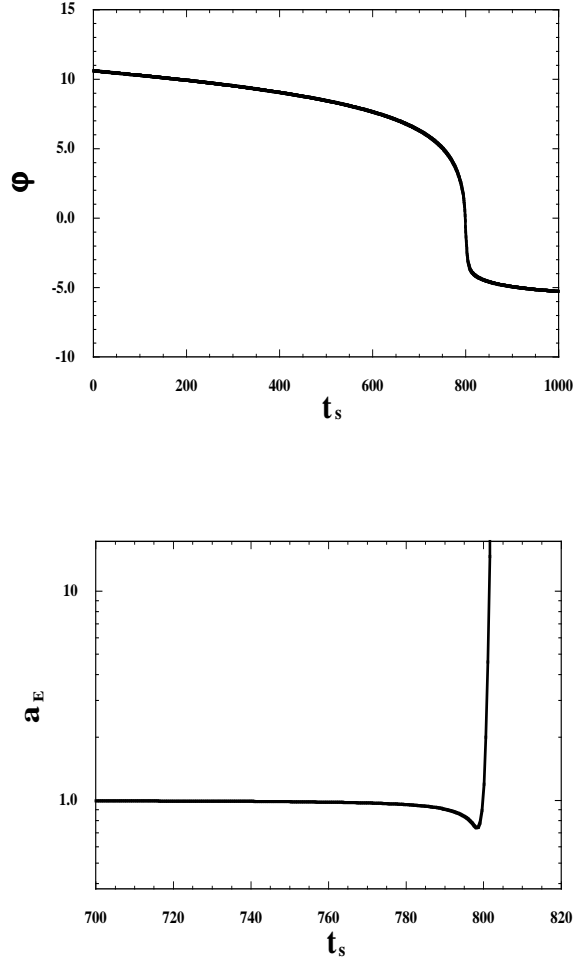}
\caption{
Evolution of $\varphi$ and $a_E$  
with $c=-1$, $d=1$, $p=0.1$, $C_1=1.0$ 
and $C_2= -1.0 \times 10^{-3}$.  In this 
case we include the correction term ${\cal L}_c$ only for $\varphi<1$.  
We choose initial conditions $\phi=-15$, $H=1.5 \times 10^{-3}$.
Prior to the collision of branes at $\varphi=0$, the universe is slowly 
contracting, which is followed by the bouncing solution 
through higher-order corrections.
}
\label{fig1}
\end{center}
\end{figure}

On the other hand, we can consider the scenario where the negative Ekpyrotic 
potential dominates initially but the higher-order correction becomes 
important when two branes approach sufficiently.  
By including the correction terms 
of ${\cal L}_c$ only for $\varphi~\lsim~1$, we have numerically 
confirmed that it is possible to obtain regular bouncing solutions, 
see Fig.~\ref{fig1}.
In this case the background 
solutions are described by Eq.~(\ref{ekysolution}) or (\ref{ekysolution2}) 
before the higher-order correction terms begin to work. 

The spectrum of scalar perturbations was studied by 
a number of authors in the cases of PBB 
cosmology \cite{Muka,CEW,Jai} and Ekpyrotic cosmology 
\cite{Lyth02,ekyper1,ekyper2,ekyper3} (see also Ref.~\cite{Wands98}).
A perturbed space-time
metric has the following form for scalar perturbations 
in an arbitrary gauge \cite{met}:
\begin{eqnarray}
\rd s^2 = -(1+2A)\rd t^2 + 2a(t)B_{,i} \rd x^i {\rm d}t 
+a^2(t)[(1-2\psi)\delta_{ij}+2E_{,i,j}] 
\rd x^i \rd x^j\,, 
\label{pmetric}
\end{eqnarray}
where a comma denotes the usual flat space 
coordinate derivative. 
The curvature perturbation, ${\cal R}$, in the comoving gauge
is given by \cite{Lyth85} 
\begin{eqnarray}
{\cal R} \equiv \psi+\frac{H}{\dot{\phi}}\delta \phi\,.
\label{metric}
\end{eqnarray}
The power spectrum of ${\cal R}$ is defined by 
${\cal P}_{\cal R} \equiv k^3|{\cal R}|^2/(2\pi^2) \propto k^{n_{\cal 
R}-1}$, where $k$ is a comoving wavenumber.
The spectral index $n_{\cal R}$ of curvature
perturbations generated before the bounce
is given by \cite{Wands98,Lyth02} 
(see also Refs.~\cite{ekyper1,ekyper2,ekyper3}):
\begin{eqnarray}
\label{speekp}
n_{\cal{R}}-1 &=& \frac{2}{1-p}\,\,\,\,\,\,\,\,\,({\rm for}~~0<p \le 1/3)\,, \\
&=& \frac{4-6p}{1-p}\,\,\,\,\,\,\,\,\,
({\rm for}~~1/3 \le p <1)\,.
\label{tilt_po}
\end{eqnarray}

We see that a scale-invariant spectrum with $n_\R=1$ is obtained
either as $p\to\infty$ in an expanding universe, corresponding to
conventional slow-roll inflation, or for $p=2/3$ during collapse
\cite{Star79,Wands98}.
In the case of the PBB cosmology ($p=1/3$) one has
$n_{\cal{R}}=4$, which is a highly blue-tilted spectrum.
The ekpyrotic scenario corresponds to a slow contraction
($0<p \ll 1$), in which case we have $n_{\cal{R}} \simeq 3$.

The spectrum (\ref{speekp}) corresponds to the one
generated before the bounce.
In order to obtain the final power spectrum at sufficient late-times
in an expanding branch, we need to connect the contracting
branch with the Friedmann (expanding) one.
As we mentioned, the two branches are joined each other
by including the corrections given in Eq.~(\ref{xifunction}).
This then allows the
study of the evolution of cosmological perturbations without having
to use matching prescriptions. The effects of the higher-order
string corrections to 
the action on the evolution of fluctuations in
the PBB cosmology was investigated
numerically in \cite{Cartier01,TBF02}. It was found
that the final spectrum of fluctuations is highly blue-tilted
($n_{\cal R} \simeq 4$)  and the result
obtained is the same as what follows from the
analysis using matching conditions between two Einstein Universes
\cite{Bru94,Der95} joined along a constant scalar field
hypersurface.

It was shown in Ref.~\cite{TBF02} that the spectrum of
curvature perturbations long after the bounce is given as
$n_{\cal R} \simeq 3$ for $0<p \ll 1$ by numerically
solving perturbation equations in a nonsingular background
regularized by the correction term (\ref{lagalpha}).
In particular comoving curvature perturbations are conserved
on cosmologically relevant scales much larger than
the Hubble radius around the bounce, which means that
the spectrum (\ref{speekp}) can be used
in an expanding background long after the bounce.

The authors in \cite{ekpyr2} showed that the spectrum of the
gravitational potential $\Phi$, 
generated before the bounce is nearly scale-invariant for $0<p \ll 1$,
i.e., $n_\Phi -1=-2p/(1-p)$.  A number of authors argued
\cite{Lyth02,ekyper1,ekyper2,ekyper3} that this corresponds
to the growing mode in the contracting phase but to the decaying mode
in the expanding phase.  
Cartier \cite{Cartier04} recently performed detailed numerical
analysis using nonsingular perturbation equations and found that in
the case of the $\alpha'$-regularized bounce both $\Phi$ and ${\cal
  R}$ exhibit the highly blue-tilted spectrum (\ref{speekp}) long
after the bounce.  It was numerically shown that the dominant
mode of the gravitational potential is fully converted into the
post-bounce decaying mode.
Similar conclusions have also been reached in investigations of
perturbations in other specific non-singular models
\cite{nonsinmodels}.
Arguments can given that the comoving curvature perturbation is
conserved for adiabatic perturbations on large scales under very
general conditions \cite{Lyth02,CNZ05}.

Nevertheless we have to caution that these studies are based on
non-singular four-dimensional bounce models and in the
Ekpyrotic/Cyclic model the bounce is only non-singular in a
higher-dimensional completion of the model \cite{Tolley02}.  The
ability of the ekpyrotic/cyclic model to produce a scale-invariant
spectrum of curvature perturbations after the bounce relies on this
higher-dimensional physics being fundamentally different from
conventional four-dimensional physics, such that the growing mode of
$\Phi$ in the contracting phase does not decay after 
the bounce \cite{Tolley03}.

\section{Dark energy}
\label{dark}

In the previous section we have discussed the role of higher-order
string corrections to the tree-level action in the context of early universe.
Recent observations show that the present universe is dominated 
by dark energy responsible for an accelerated expansion.
When the universe is dominated by 
a phantom matter ($w<-1$), this leads to the growth of 
the energy and  curvature invariants.
In such a circumstance higher-order string
corrections may be important when the energy scale grows
up to a Planck one. We are interested in the effect of 
such corrections around the Big Rip singularity.
In fact the Big Rip singularity can be avoided in the presence 
of such corrections as we will see in this section.
We shall also derive cosmological solutions for an effective 
string Lagrangian together with a barotropic perfect fluid.

Our starting action is the generalization of 
(\ref{effactions}):
\be\label{act}
{\cal S}=\int {\rm d}^D x \sqrt{-g}\left[\frac12 
f(\phi,R)-\frac12\omega(\phi)(\nabla\phi)^2
-V(\phi)+\xi(\phi){\cal L}_c^{(\phi)}
+{\cal L}_m^{(\phi)} \right]\,,
\ee
where $\phi$ is a scalar field corresponding either to the dilaton or to 
another modulus, and $f$ is a generic function of the 
scalar field and the Ricci scalar $R$.
$\omega$, $\xi$ and $V$ are functions of 
$\phi$. In this section we do not consider the cosmological 
dynamics in the presence of the field potential $V$.
${\cal L}_m^{(\phi)}$ is the Lagrangian of a barotropic
perfect fluid with energy density $\rho$ and 
pressure density $p$. We assume that the barotropic 
index, $w\equiv p/\rho$, is a constant.
In general the fluid can be coupled to the scalar field $\phi$.
The $\alpha'$-order quantum corrections are encoded in the term
\be 
\label{Lc}
{\cal L}_c^{(\phi)} 
=a_1R_{\alpha\beta\mu\nu}R^{\alpha\beta\mu\nu}
+a_2R_{\mu\nu}R^{\mu\nu}
+a_3R^2+a_4(\nabla\phi)^4\,,
\ee
where $a_i$ are coefficients depending 
on the string model one is considering. 
We are most interested in the Gauss-Bonnet parametrization 
($a_1=a_3=1$, $a_2=-4$ and $a_4=-1$) discussed in the 
previous section, but we keep 
the coefficients general in deriving basic equations.

For a flat FRW background with a scale factor $a$, 
we obtain the Friedmann equation \cite{Cartier01,CTS}
\ba
\label{FReq1}
d(d-1)FH^2 = RF-f-2dH\dot{F}+2(\rho+\rho_\phi+\xi\rho_c),
\ea
where $d=D-1$ and 
\ba
\label{FReq2}
\rho_\phi &=& (\omega/2)\dot{\phi}^2+V, \\
\label{FReq3}
\rho_c &=& 3a_4\dot{\phi}^4-d[4c_1\Xi H^3+(d-3)c_1H^4+c_2(2\Xi 
H\dot{H}+2dH^2\dot{H}-\dot{H}^2+2H\ddot{H})].
\ea
Here $F\equiv \partial f/\partial R$, $\Xi\equiv \dot{\xi}/\xi$, and
\ba
c_1  \equiv 2a_1+da_2+d(d+1)a_3,\,\,\,\,\,
c_2  \equiv 4a_1+(d+1)a_2+4da_3.
\ea
In four dimensions ($d=3$), the 
coefficients read $c_1=2a_1+3a_2+12a_3$ and 
$c_2=4(a_1+a_2+3a_3)$, while the $H^4$ term in $\rho_c$ vanishes. At 
low energy it was shown that the unique higher-order gravitational 
Lagrangian giving a theory without ghosts is the Gauss--Bonnet 
one ($a_1=a_3=1$, $a_2=-4$, $a_4=-1$). In this case, $c_2$ vanishes identically 
while $c_1=2+d(d-3)$. With fixed dilaton coupling ($\Xi=0$)
equation (\ref{FReq1}) reduces to the standard Friedmann 
equation in four dimensions, 
in agreement with the fact that the GB term is topological
when $d=3$. In three dimensions ($d=2$), the GB higher-derivative 
contribution vanishes identically except for the $\dot{\phi}^4$ term.

The continuity equation for the dark energy fluid contains a source 
term given by the coupling between this fluid and the scalar 
field $\phi$. We choose the covariant coupling considered 
in \cite{Luca}: $\delta {\cal S}_m/\delta\phi
=-\sqrt{-g}\,Q(\phi)\rho$, where ${\cal S}_m=\int {\rm d}^Dx 
\sqrt{-g} {\cal L}_m^{(\phi)}$ and $Q(\phi)$ is an unknown 
function which we shall set to a constant later. 
In synchronous gauge we have
\be\label{drho}
\dot{\rho}+dH\rho(1+w)=Q\dot{\phi}\rho,
\ee
while the equation of motion for the field $\phi$ is
\be\label{pheom}
\omega(\ddot{\phi}+dH\dot{\phi})+V'-\xi'{\cal 
L}_c^{(\phi)}+4a_4\xi\dot{\phi}^2(3\ddot{\phi}
+dH\dot{\phi}+\Xi\dot{\phi})
+\left(\dot{\omega}\dot{\phi}-\omega'\frac{\dot{\phi}^2}{2}
-\frac{f'}{2}\right)=-Q\rho,
\ee
where the Lagrangian of the quantum correction 
is written as
\be
{\cal L}_c^{(\phi)} = 
d\left[(d+1)c_1H^4+4c_1H^2\dot{H}+c_2\dot{H}^2\right]
+a_4\dot{\phi}^4.
\ee
Equations (\ref{FReq1})--(\ref{pheom}) are the master 
equations of the physical system under study.
We note that the $\dot{\phi}^4$ term can be important 
even in the absence of curvature corrections
in a dilatonic ghost condensate model \cite{PT04}.

The massless dilaton discussed 
in the previous section corresponds to
\ba
F = -\omega =e^{-\phi},\,\,\,\,\,
V = 0,\,\,\,\,\,
\xi = \frac{\lambda}{2} e^{-\phi}.
\label{tree}
\ea
The full contribution of $n$-loop corrections is 
given by Eq.~(\ref{xifunction}). In this work we shall take only 
the tree-level term (\ref{tree}) into account. 

Generally moduli fields appear whenever a submanifold of the target 
spacetime is compactified with compactification radii described by the expectation 
values of the moduli themselves. In the case of a single modulus (one common 
characteristic length) and heterotic string ($\lambda=1/8$), 
the four-dimensional action corresponds to \cite{ART}
\ba
\label{modu1}
F = 1,\,\,\,\,
\omega = 3/2,\,\,\,\,
a_4 = 0,\,\,\,\,
\xi = -\frac{\delta}{16}
\ln [2e^\phi\eta^4(ie^\phi)],
\label{modu2}
\ea
where $\eta$ is the Dedekind function and $\delta$ is a constant proportional 
to the 4D trace anomaly. $\delta$ depends on the number of chiral, vector, 
and spin-$3/2$ massless supermultiplets of the $N=2$ sector of the theory. 
In general it can be either positive or negative, 
but it is positive for the theories in which not
too many vector bosons are present. 
Again the scalar field corresponds to a flat direction 
in the space of nonequivalent vacua and $V=0$.
At large $\phi$ the last equation can be approximated as
\ba
\xi \approx \xi_0 \cosh\phi,\,\,\,\,\,
\xi_0 \equiv \frac{\pi \delta}{24},\label{ximodu2}
\ea
which we shall use instead of the exact expression. 
In fact it was shown in Ref.~\cite{YMO} that 
this approximation gives results very close to those of the exact case.

\subsection{Modulus driven solution}
\label{modu}

In Ref.~\cite{CTS} cosmological solutions based on the action (\ref{act})
without a potential ($V=0$) were discussed in details for three cases--(i) 
fixed scalar field ($\dot{\phi}=0$), (ii)
linear dilaton ($\dot{\phi}={\rm const}$), and (iii) 
logarithmic modulus ($\dot{\phi} \propto 1/t$).
In the case (i) we obtain geometrical inflationary solutions only for $D \ne 4$.
In the case (ii) pure de-Sitter solutions exist in string frame, but 
this corresponds to a contracting universe in Einstein frame.
These solutions are not realistic when we apply to dark energy.
In what follows we shall focus on cosmological solutions 
in the case (iii) with a fixed dilaton.

Introducing the following new variables 
\be
x\equiv H\,,\qquad  y\equiv\dot{H}\,,\qquad 
u\equiv\phi\,,\qquad
v\equiv\dot{\phi}\,,\qquad z\equiv\rho\,,
\ee
the equations of motion for the modulus action corresponding 
to Eqs.~(\ref{FReq1})-(\ref{pheom}) read
\ba 
& & \dot{x} = y,\label{modeq1}\\ 
& & \dot{y} = \frac{2z+3v^2/2-d(d-1)x^2-8dc_1\dot{\xi} x^3-2d(d-3)c_1\xi x^4
-2d c_2\xi y(2\Xi x+2dx^2-y)}{4dc_2\xi x}, \nonumber \\ \label{modeq2}\\
& & \dot{v} = \frac{2d}{3}\xi'[(d+1)c_1 x^4+4 c_1 x^2 y+c_2 y^2]-dxv
-\frac{2Q}{3}z,\label{modeq3}\\
& & \dot{z} = \left[-dx(1+w)+Qv \right]z.
\label{modeq4}
\ea
While only derivatives of $\xi$ appear in the equations of motion 
for $d=3$ and $c_2=0$ (GB case), there are non-vanishing 
contributions of $\xi$ itself for general coefficients $c_i$.
When $c_2=0$, the equations of motion for $x$ and $v$ read
\ba
\dot{x} &=& \frac{\dot{z}+3v\dot{v}/2-dc_1x^3[4\ddot{\xi}+(d-3)\dot{\xi}x]}
{dx [(d-1)+12c_1\dot{\xi}x+4(d-3)c_1\xi x^2]},
\label{eqc201}\\
\frac{\dot{v}}{v^2} &=& \frac{2d}{3}c_1\xi'\frac{x^2}{v^2}
[(d+1)x^2+4\dot{x}]-d\frac{x}{v}-\frac{2Q}{3}\frac{z}{v^2},
\label{eqc202}
\ea
while the Friedmann equation is
\be\label{femo}
d(d-1)\frac{x^2}{v^2}-\frac32-2\frac{z}{v^2}+2dc_1\xi \frac{x^3}{v^2}[4\Xi+(d-3)x]=0.
\ee
In addition $c_1$ can be set equal to 1 and absorbed in the definition of $\xi_0$, 
so that the coefficient $c_2$ is the only free parameter of the higher-order Lagrangian.

We search for future asymptotic solution of the form
\ba
\label{asso1}
x &\sim& \omega_1 t^\beta,\quad y \sim \beta\omega_1 t^{\beta-1}, \quad
u \sim  u_0+\omega_2\ln t,\quad v \sim \frac{\omega_2}{t},\quad
\xi \sim \frac12 \xi_0e^{u_0}t^{\omega_2},\\
z &\sim& z_0t^{Q\omega_2}\exp\left[-\frac{d(1+w)\omega_1}
{\beta+1}t^{\beta+1}\right],\qquad \beta\neq -1,\label{zne1}\\
z &\sim& z_0t^{\alpha},\qquad \beta= -1,\label{asso5}
\ea
where the barotropic index $w$ is constant and
\be\label{Qw}
\alpha \equiv Q\omega_2-d(1+w)\omega_1.
\ee
We define $\tilde{\delta}\equiv (1/2)c_1\xi_0e^{u_0}$,
so that in any claim involving the sign of $\tilde{\delta}$ ($\delta$) a positive $c_1$ 
coefficient is understood. In order to find a solution in the limit $t \to +\infty$, 
one has to match the exponents of $t$ to get algebraic equations 
in the parameters $\beta, \omega_i,c_2$ and $\tilde{\delta}$. 

In Ref.~\cite{ART,CTS} the following four solutions were found.
We can obtain a number of analytic solutions depending on the regimes
we are in:
\begin{enumerate}
\item 
A low-curvature regime in which $\xi$ terms 
are subdominant at late times.
\item 
An intermediate regime 
where some terms in the equations of motion, either coupled to $\xi$ or not,
are damped. 
\item 
A high-curvature regime in which $\xi$ terms dominate.
\item 
A solution of the form (\ref{asso1})--(\ref{asso5}) 
for the full equations of motion.
\end{enumerate}
Below we summarize the properties of each solution.
\begin{enumerate}

 \item \underline{A low-curvature regime}

In this regime the solution is given by 
\ba \label{grsol}
\beta=-1,\qquad \omega_2<2,
\ea
with the constraints
\ba
\omega_1 = \frac{1}{d}-\frac{2Qz_0t^{\alpha+2}}{3d\omega_2}, \quad
3\omega_2^2 = 2d(d-1)\omega_1^2-4z_0t^{\alpha+2},\quad
\alpha \leq -2.\label{Qeq}
\ea

\item \underline{An intermediate regime}

In this regime the solution exists only for $c_2=0$ and is given by 
\ba
\label{minsol}
\beta=-2,\quad \omega_2=5,\quad Q\leq-2/5,\quad \omega_1^3
=\frac{1}{16d\tilde{\delta}}\left(15-2Qz_0t^{5Q+2}\right),
\ea
for a non-vanishing fluid.
The condition, $\tilde{\delta}>0$, is required in order to obtain 
an expanding solution characterized by $a(t)\sim a_0\exp(-\omega_1/t)$. 
This solution reaches Minkowski spacetime
asymptotically. If the fluid decays, then one recovers the $C_\infty$ solution 
of Ref.~\cite{ART} with $d=3$ and $z=0$. 

\item \underline{A high-curvature regime}

In this regime the solution is given by 
\ba
\label{rem}
\beta=-1,\quad \omega_2>2,\quad \alpha\leq \omega_2-4,
\ea
together with constraint equations
\ba
&& \tilde{\delta}d\omega_1^2[(d+1)\omega_1^2-4\omega_1+c_2]
-Qz_0t^{\alpha-\omega_2+4}=0,\label{rem1}\\
&& d\tilde{\delta}\omega_1^2 [(3-d)\omega_1^2-4\omega_2\omega_1
+c_2(2d\omega_1+2\omega_2-3)]+z_0t^{\alpha-\omega_2+4}=0.\label{rem2}
\ea
In the GB case ($c_2=0$) the solution corresponding to a decaying 
fluid ($\alpha<\omega_{2}-4$) is 
\ba
\omega_1 = \frac{4}{d+1},\quad 
\omega_2 = \frac{3-d}{d+1},
\label{remsol}
\ea
which contradicts the condition (\ref{rem}) in any dimension.
Hence in the GB scenario with $\omega_2\neq 4$ 
only the marginal case $\alpha=\omega_2-4$ is allowed.

\item \underline{An exact solution}

An exact solution which is valid at all times is
\ba
\label{exsol}
\beta=-1,\quad \omega_2=2,
\ea
together with the constraints on $\omega_1$:
\ba
& & 2\tilde{\delta} d\omega_1^2[(d+1)\omega_1^2-4\omega_1+c_2]
-6d\omega_1+6-Qz_0t^{\alpha+2}=0,\label{exso1}\\
& & 2\tilde{\delta} d\omega_1^2[(d-3)\omega_1^2+8\omega_1
-c_2(2d\omega_1+1)]+d(d-1)\omega_1^2-6-2z_0t^{\alpha+2}=0,\label{exso2}
\ea
and Eq.~(\ref{Qeq}). 

\end{enumerate}

The low-curvature solution (\ref{grsol}) and the Minkowski solution
(\ref{minsol}) can be joined each other
if the coupling constant $\delta$ given 
in Eq.~(\ref{modu2}) is negative \cite{ART,Kanti}.
The exact solution (\ref{exsol}) is found to be unstable
in numerical simulations of Ref.~\cite{CTS}.
In the asymptotic future the solutions tend to approach the 
low-curvature one given by Eq.~(\ref{grsol}) rather than 
the others, irrespective of the sign of 
the modulus-to-curvature coupling $\delta$.

\subsection{Constraints from the recent universe}

We compare the observational constraints on $\omega_1$ 
for the recent evolution of the universe with the modulus solutions found 
in the previous section ($d=3$).
The situation we study is the case in which a
perfect fluid is vanishing asymptotically.
The results are summarized in Table \ref{tab1} 
at the 68\% confidence level. 
We also address the cases in the presence of 
a cosmological constant $\Lambda$.
Note that the solution (\ref{minsol})
in an intermediate regime is discarded.

\begin{vchtable}[htb]
\begin{center}
\begin{tabular}{l|c|c}\hline
Solutions  $t\to \infty$ & $\omega_1^{(0)}$   & Stability \\ 
\hline
Low curvature &                	& Yes       \\ \hline
High curvature, $c_2=0$    &           & $\Lambda>0$ \\ \hline
High curvature, $c_2\neq 0$, $\Lambda=0$    &              								   &          \\
High curvature, $c_2\neq 0$, $\Lambda\neq0$ & $\omega_1^+$, 
$c_2<-8$ & $\omega_1^+$, $\Lambda>0$ \\  \hline
Exact, $c_2=0$                              &                                
&\\ \hline
Exact, $c_2 \ne 0$ &     Yes             & \\ 
\hline
\end{tabular}
\caption{\label{tab1} 
Constraints on modulus solutions 
in the asymptotic future for the Hubble parameter $H=\omega_1 t^{-1}$, 
vanishing fluid, and $d=3$. 
Blank entries are excluded by experiments or numerical analysis.}
\end{center}
\end{vchtable}

The logarithmic modulus solution with the GB parametrization and no extra fluid does not provide 
a viable cosmological evolution in the current universe. 
In the next subsection, however, we will see 
that the GB case in the presence of dark energy fluid may exhibit 
interesting features for the future evolution of the universe.
The low redshift constraint on $c_2$ 
for the high curvature solution ($\Lambda \neq 0$) 
can be relaxed up to $c_2<-1$ at the 99\% confidence level.
Hence we have shown that there are models which can in principle 
explain the current acceleration 
without using the dark energy fluid.

The situation becomes more complicated 
in the presence of a barotropic fluid.
The low-curvature solution can describe the very recent universe 
if $Q$ is negative and non-vanishing. The other cases crucially 
depend upon the interplay between all the theoretical parameters.

\subsection{Dark energy universe with modulus gravity}

In the universe dominated by a phantom fluid ($w<-1$), the 
energy density of the universe
continues to grow and the Hubble rate eventually exhibits 
a divergence at finite time (Big Rip).
Then the effect of higher-order curvature corrections
can become important when the energy density grows up to the 
Planck scale. In fact it was shown that quantum curvature corrections coming from 
conformal anomaly can moderate the future singularities \cite{NO,NOT}.

We would like to consider the effect of $\alpha'$
quantum corrections when the curvature of the universe increases
in the presence of a phantom fluid. We shall concentrate on the modulus case 
with $\xi$ given by Eq.~(\ref{ximodu2}).
Our main interest is the cosmological evolution in four 
dimensions ($d=3$) in the presence of a GB term.
The dilaton is assumed to be fixed, 
so that there are no long-range forces to take into account except gravity.

{}From the discussion in subsection \ref{modu}, the growth of 
the barotropic fluid is weaker than that of the Hubble rate 
when the condition
\ba
\alpha=Q\omega_{2}-d(1+w)\omega_{1} \le -2\,,
\ea
is satisfied.
This condition is not achieved for a phantom fluid 
when the coupling $Q$ between the fluid and the field $\phi$ 
is absent ($Q=0$).

\begin{figure}[htb]
\begin{center}
\includegraphics[width=8cm, height=8cm]{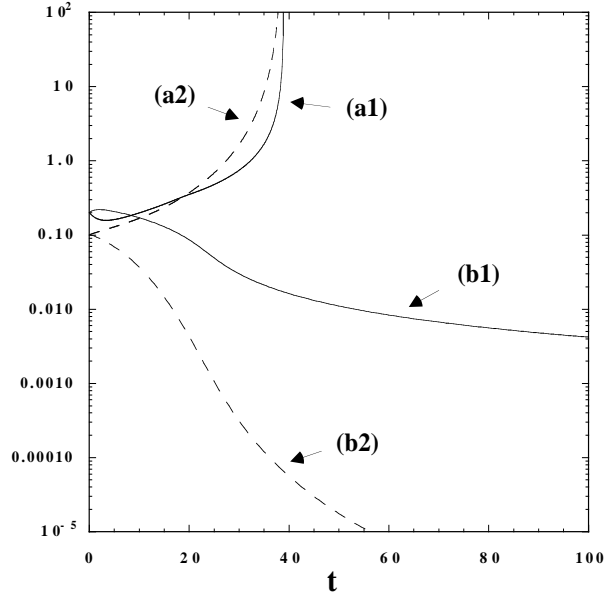}
\caption{
Evolution of $H$ and $\rho$ with $\xi_{0}=-2$, $w=-1.1$ for (a) 
$Q=0$ and (b) $Q=-5$. We choose initial conditions as $H_{i}=0.2$, 
$\phi_{i}=2.0$ and $\rho_{i}=0.1$.
The curves (a1) and (b1) represent the evolution of $H$ for
$Q=0$ and $Q=-5$, respectively, while the curves (a2) and (b2)
show the evolution of $\rho$ for corresponding $Q$.}
\label{fig2}
\end{center}
\end{figure}

In Ref.\,\cite{CTS} the equations of motion were solved numerically 
by varying initial conditions of $H$, $\phi$ and $\rho$.
When $\delta<0$, we numerically found that 
the solutions approach a Big Rip singularity for $Q=0$
and $w<-1$ (see Fig.~\ref{fig2}).
The condition (\ref{Qeq}) can be satisfied for negative $Q$
provided that $\omega_{2}$ is positive.
In Fig.~\ref{fig2} we plot the evolution of $H$ and $\rho$
for $Q=-5$ and $w=-1.1$. In this case $\rho$ decreases 
faster than $t^{-2}$, which means that 
the energy density of the fluid eventually becomes negligible 
relative to that of the modulus. 
Hence the universe approaches the low-curvature
solution given by Eq.~(\ref{grsol}) at late times, thereby 
showing the avoidance of Big Rip singularity even for $w<-1$.
By substituting the asymptotic values $\omega_{1}=1/3$ and 
$\omega_{2}=2/3$ in equation~(\ref{Qeq}), the condition for 
decaying fluid reads $Q<3(w-1)/2=-3.15$.
We checked that the Big Rip singularity 
can be avoided in a wide range of the parameter space  
for negative $Q$. These results do 
not change even for smaller values of $\delta$ such as 
$|\delta|={\cal O}(1)$ (corresponding to $|\xi_0| = 0.1$).

When $Q$ is positive, the condition (\ref{Qeq}) is not satisfied
for $\omega_{2}>0$. However numerical calculations show that 
$\dot{\phi}$ becomes negative even if $\dot{\phi}>0$ initially.
We found that the system approaches the low-energy regime  
characterized by $\omega_{1}=1/3$ and $\omega_{2}=-2/3$.
Since $\omega_{2}<0$, the Big Rip
singularity can be avoided even for positive $Q$.
In fact we numerically checked that the Hubble rate 
continues to decrease provided that the condition (\ref{Qeq})
is satisfied in the asymptotic regime.

When $\delta>0$, there is another interesting situation in 
which the Hubble rate decreases in spite of the increase of the 
energy density of the fluid. This corresponds to the solution in the 
high-curvature regime in which the growing energy density $\rho$
can balance with the GB term ($\rho \approx 24H^3\dot{\xi}$ in 
the Friedmann equation). In this case the Big Rip does not appear 
even when $w<-1$ and $Q=0$. Thus the GB corrections 
provide us several interesting possibilities to avoid 
the Big Rip singularity.

\subsection{Other corrections}

We should mention several attempts to apply 
curvature corrections to dark energy.
Nojiri {\it et al.} \cite{NOS} studied the case of the tree-level 
dilatonic-type coupling (\ref{tree}) in the presence of 
an exponential potential and showed that the scalar-Gauss-Bonnet
coupling acts against the occurrence of the Big Rip singularity.
In Ref.~\cite{Topo} the evolution of phantom dark energy 
universe was studied with fixed dilaton and modulus
by taking into account string corrections up to the 4-th order. 
It was found that the universe reaches a Big Crunch singularity 
instead of a Big Rip one for type II strings, while the 
Big Rip singularity is not avoided for heterotic and 
bosonic strings.

Apart from string corrections, a number of authors 
\cite{Eli,NO,NOT,Sri,Tre} studied the 
effect of quantum backreactions of conformal matter
around several singularities which can appear in future.
They usually contain second-order curvature
corrections such as the Gauss-Bonnet term and the square of a Weyl 
tensor. In Ref.~\cite{NOT} it was shown that quantum corrections
coming from conformal anomaly can be important when the curvature of 
the universe grows large, which typically moderates
future singularities. Finally we note that loop quantum cosmology 
leads to a modified Friedmann equation when the energy scale
grows to a Planck one, which typically gives us a regular 
cosmological evolution without future singularities \cite{SST06}.

\section{Conclusions}

In this article we have discussed cosmological implications of 
higher-order string corrections to the tree-level effective action.
In the context of Pre-Big-Bang (PBB) and Ekpyrotic cosmologies
regular bouncing solutions can be constructed by including 
such corrections.
This allows us to evaluate the spectrum of
density perturbations long after the bounce.
For the correction terms given by Eq.~(\ref{xifunction}) 
we found that the spectra of scalar perturbations are highly 
blue-tilted: $n_{\cal R}=4$ in the PBB case and 
$n_{\cal R}=3$ in the Ekpyrotic case.
This is different from the nearly scale-invariant spectrum 
($n_{\cal R} \simeq 1$) observed in CMB anisotropies.
As long as nonsingular bouncing solutions are constructed
by using the correction terms presented in this paper, 
we need another scalar field 
(e.g., curvaton \cite{curvaton}) to generate nearly 
scale-invariant density perturbations.

We also applied second-order string corrections to 
dark energy.
In particular we reviewed several cosmological solutions
in the presence of modulus-type corrections with a fixed dilaton.
In the asymptotic future the solutions tend to approach the 
low-curvature one given by equation (\ref{grsol}) rather than 
the others, irrespective of the sign of the modulus-to-curvature coupling $\delta$.
We placed constraints on the viability of modulus-driven 
solutions using the current observational data.
The Gauss-Bonnet parametrization is excluded 
in any of the above mentioned regimes when a barotropic fluid 
is vanishing; see table \ref{tab1}.
In the presence of a phantom dark energy fluid we discussed
the effect of the modulus coupling with 
Gauss-Bonnet curvature invariants.
It is possible to consider 
a situation in which the energy density of the fluid decays 
when the coupling $Q$ between the field $\phi$
and the phantom fluid ($w<-1$) is present.
We showed that the Big Rip singularity can be avoided for the coupling $Q$
which satisfies the condition $Q\omega_{2}-3(1+w)\omega_{1}<-2$
asymptotically. This is actually achieved 
irrespective of the sign of $Q$ and the asymptotic solutions are 
described by the low-curvature one given by Eq.~(\ref{grsol}).
We also briefly mentioned the effect of other forms of 
higher-order string corrections to future singularities.

Thus we showed that string corrections
can be important in a number of cosmological situations.
We hope that the development of string theory will further 
provide us rich and fruitful implications to cosmology.

\section*{ACKNOWLEDGMENTS}
The author thanks the organizers of ``Pomeranian Workshop 
in Fundamental Cosmology'',
especially Mariusz Dabrowski, for supporting visit to the 
wonderful workshop. 
It is also a pleasure to thank my collaborators 
for fruitful discussions.
This work is supported by JSPS
(Grant No.\,30318802).



\begin{thebibliography}{10}
    
\bibitem{Lidsey99}
J.~E.~Lidsey, D.~Wands and E.~J.~Copeland,
Phys.\ Rept.\  {\bf 337}, 343 (2000).

\bibitem{Mariusz}
M.~P.~Dabrowski,
Annalen Phys.\  {\bf 10} (2001) 195.

\bibitem{Gasperini02}
M.~Gasperini and G.~Veneziano,
Phys.\ Rept.\  {\bf 373}, 1 (2003).

\bibitem{Veneziano91}
G.~Veneziano,
Phys.\ Lett.\ B {\bf 265}, 287 (1991); 
K.~A.~Meissner and G.~Veneziano,
Phys.\ Lett.\ B {\bf 267}, 33 (1991).

\bibitem{pbb}
M.~Gasperini and G.~Veneziano,
Astropart.\ Phys.\  {\bf 1} (1993) 317.

\bibitem{ekpyr}
J.~Khoury, B.~A.~Ovrut, P.~J.~Steinhardt and N.~Turok,
Phys.\ Rev.\ D {\bf 64}, 123522 (2001).

\bibitem{ekpyr2}
J.~Khoury, B.~A.~Ovrut, P.~J.~Steinhardt and N.~Turok,
Phys.\ Rev.\ D {\bf 66}, 046005 (2002).

\bibitem{cyclic}
P.~J.~Steinhardt and N.~Turok,
Science {\bf 296}, 1436 (2002);
Phys.\ Rev.\ D {\bf 65}, 126003 (2002).
  
\bibitem{Gas96}
M.~Gasperini, M.~Maggiore and G.~Veneziano,
Nucl.\ Phys.\ B {\bf 494}, 315 (1997).

\bibitem{Bru}
R.~Brustein and R.~Madden,
Phys.\ Rev.\ D {\bf 57}, 712 (1998).

\bibitem{TBF02}
S.~Tsujikawa, R.~Brandenberger and F.~Finelli,
Phys.\ Rev.\ D {\bf 66}, 083513 (2002).

\bibitem{Anto93}
I.~Antoniadis, J.~Rizos and K.~Tamvakis,
Nucl.\ Phys.\ B {\bf 415}, 497 (1994).

\bibitem{WMAP}
D.~N.~Spergel {\it et al.},
Astrophys.\ J.\ Suppl.\  {\bf 148}, 175 (2003).

\bibitem{inflationworks}
A.~A.~Starobinsky,
Phys.\ Lett.\ B {\bf 91}, 99 (1980);
M.~C.~Bento and O.~Bertolami,
Phys.\ Lett.\ B {\bf 228}, 348 (1989);
Class.\ Quant.\ Grav.\  {\bf 17}, 4855 (2000);
S.~W.~Hawking, T.~Hertog and H.~S.~Reall,
Phys.\ Rev.\ D {\bf 63}, 083504 (2001);
K.~i.~Maeda and N.~Ohta,
Phys.\ Lett.\ B {\bf 597}, 400 (2004);
Phys.\ Rev.\ D {\bf 71}, 063520 (2005);
K.~Akune, K.~i.~Maeda and N.~Ohta,
Phys.\ Rev.\ D {\bf 73}, 103506 (2006).

\bibitem{SN}
A.~G.~Riess {\it et al.}  [Supernova Search Team Collaboration],
Astron.\ J.\  {\bf 116}, 1009 (1998);
S.~Perlmutter {\it et al.},
Astrophys.\ J.\  {\bf 517}, 565 (1999);
A.~G.~Riess {\it et al.},
Astrophys.\ J.\  {\bf 607}, 665 (2004).

\bibitem{obser}
U.~Alam, V.~Sahni, T.~D.~Saini and A.~A.~Starobinsky,
Mon.\ Not.\ Roy.\ Astron.\ Soc.\  {\bf 354}, 275 (2004);
P.~S.~Corasaniti, M.~Kunz, D.~Parkinson, E.~J.~Copeland
and B.~A.~Bassett,
Phys.\ Rev.\ D {\bf 70}, 083006 (2004).

\bibitem{reviewdark}
V.~Sahni and A.~A.~Starobinsky, Int.\ J.\ Mod.\ Phys.\ D \textbf{9},
373 (2000); V.~Sahni, Lect.\ Notes Phys.\ {} \textbf{653}, 141 (2004);
S.~M.~Carroll, Living Rev.\ Rel.\ {} \textbf{4}, 1 (2001);
T.~Padmanabhan, Phys.\ Rept.\ {} \textbf{380}, 235 (2003);
P.~J.~E.~Peebles and B.~Ratra, Rev.\ Mod.\
Phys.\ {} \textbf{75}, 559 (2003);
E.~J.~Copeland, M.~Sami and S.~Tsujikawa, arXiv:hep-th/0603057.

\bibitem{Caldwell}
R.~R.~Caldwell,
Phys.\ Lett.\ B {\bf 545}, 23 (2002).

\bibitem{CKW}
R.~R.~Caldwell, M.~Kamionkowski and N.~N.~Weinberg,
Phys.\ Rev.\ Lett.\  {\bf 91}, 071301 (2003).

\bibitem{Carroll}
S.~M.~Carroll, M.~Hoffman and M.~Trodden,
Phys.\ Rev.\ D {\bf 68}, 023509 (2003).

\bibitem{Singh}
P.~Singh, M.~Sami and N.~Dadhich,
Phys.\ Rev.\ D {\bf 68}, 023522 (2003).

\bibitem{NOT}
S.~Nojiri, S.~D.~Odintsov and S.~Tsujikawa,
Phys.\ Rev.\ D {\bf 71}, 063004 (2005).

\bibitem{NOS}
S.~Nojiri, S.~D.~Odintsov and M.~Sasaki,
 Phys.\ Rev.\ D {\bf 71}, 123509 (2005).

\bibitem{Topo}
M.~Sami, A.~Toporensky, P.~V.~Tretjakov and S.~Tsujikawa,
Phys.\ Lett.\ B {\bf 619}, 193 (2005).
 
\bibitem{CTS}
G.~Calcagni, S.~Tsujikawa and M.~Sami,
Class.\ Quant.\ Grav.\  {\bf 22}, 3977 (2005).

\bibitem{ACD}
L.~Amendola, C.~Charmousis and S.~C.~Davis,
arXiv:hep-th/0506137.

\bibitem{Neu}
B.~M.~N.~Carter and I.~P.~Neupane,
arXiv:hep-th/0510109;
arXiv:hep-th/0512262;
I.~P.~Neupane,
arXiv:hep-th/0602097.

\bibitem{Cog}
G.~Cognola, E.~Elizalde, S.~Nojiri, S.~D.~Odintsov and S.~Zerbini,
Phys.\ Rev.\ D {\bf 73}, 084007 (2006).

\bibitem{Cal06}
G.~Calcagni, B.~de Carlos and A.~De Felice,
arXiv:hep-th/0604201.

\bibitem{Are}
I.~Y.~Aref'eva and A.~S.~Koshelev,
arXiv:hep-th/0605085.

\bibitem{NOS2}
S.~Nojiri, S.~D.~Odintsov and M.~Sami,
arXiv:hep-th/0605039.

\bibitem{KM06}
T.~Koivisto and D.~F.~Mota,
arXiv:astro-ph/0606078.

\bibitem{Foffa99}
S.~Foffa, M.~Maggiore and R.~Sturani,
Nucl.\ Phys.\ B {\bf 552}, 395 (1999).

\bibitem{Cartier99}
C.~Cartier, E.~J.~Copeland and R.~Madden,
JHEP {\bf 0001}, 035 (2000).

\bibitem{Durrer02}
R.~Durrer and F.~Vernizzi,
Phys.\ Rev.\ D {\bf 66}, 083503 (2002).

\bibitem{LM85}
F.~Lucchin and S.~Matarrese,
Phys.\ Rev.\ D {\bf 32}, 1316 (1985).

\bibitem{Met}
R.~R.~Metsaev and A.~A.~Tseytlin,
Nucl.\ Phys.\ B {\bf 293}, 385 (1987).

\bibitem{Muka}
R.~Brustein, M.~Gasperini, M.~Giovannini, V.~F.~Mukhanov and G.~Veneziano,
Phys.\ Rev.\ D {\bf 51}, 6744 (1995).

\bibitem{CEW}
E.~J.~Copeland, R.~Easther and D.~Wands,
Phys.\ Rev.\ D {\bf 56}, 874 (1997).

\bibitem{Jai}
J.~c.~Hwang,
Astropart.\ Phys.\  {\bf 8}, 201 (1998).

\bibitem{Lyth02}
D.~H.~Lyth,
Phys.\ Lett.\ B {\bf 524}, 1 (2002);
Phys.\ Lett.\ B {\bf 526}, 173 (2002).

\bibitem{ekyper1}
R.~Brandenberger and F.~Finelli,
JHEP {\bf 0111}, 056 (2001).

\bibitem{ekyper2}
J.~c.~Hwang,
Phys.\ Rev.\ D {\bf 65}, 063514 (2002).

\bibitem{ekyper3}
S.~Tsujikawa,
Phys.\ Lett.\ B {\bf 526}, 179 (2002).

\bibitem{Wands98}
D.~Wands,
Phys.\ Rev.\ D {\bf 60}, 023507 (1999).

\bibitem{met}
H.~Kodama and M.~Sasaki,
Prog.\ Theor.\ Phys.\ Suppl.\  {\bf 78}, 1 (1984);
V.~F.~Mukhanov, H.~A.~Feldman and R.~H.~Brandenberger,
Phys.\ Rept.\  {\bf 215}, 203 (1992);
B.~A.~Bassett, S.~Tsujikawa and D.~Wands,
Rev.\ Mod.\ Phys.\  {\bf 78}, 537 (2006).

\bibitem{Lyth85}
D.~H.~Lyth,
Phys.\ Rev.\ D {\bf 31}, 1792 (1985).

\bibitem{Star79}
A.~A.~Starobinsky,
JETP Lett.\  {\bf 30} (1979) 682
[Pisma Zh.\ Eksp.\ Teor.\ Fiz.\  {\bf 30} (1979) 719].

\bibitem{Cartier01}
C.~Cartier, J.~c.~Hwang and E.~J.~Copeland,
Phys.\ Rev.\ D {\bf 64}, 103504 (2001).

\bibitem{Bru94}
R.~Brustein, M.~Gasperini, M.~Giovannini, 
V.~F.~Mukhanov and G.~Veneziano,
Phys.\ Rev.\ D {\bf 51}, 6744 (1995).

\bibitem{Der95}
N.~Deruelle and V.~F.~Mukhanov,
Phys.\ Rev.\ D {\bf 52}, 5549 (1995).

\bibitem{Cartier04}
C.~Cartier,
arXiv:hep-th/0401036.

\bibitem{nonsinmodels}
M.~Gasperini, M.~Giovannini and G.~Veneziano,
Phys.\ Lett.\ B {\bf 569}, 113 (2003);
M.~Gasperini, M.~Giovannini and G.~Veneziano,
Nucl.\ Phys.\ B {\bf 694}, 206 (2004);
L.~E.~Allen and D.~Wands,
Phys.\ Rev.\ D {\bf 70}, 063515 (2004);
T.~J.~Battefeld and G.~Geshnizjani,
Phys.\ Rev.\ D {\bf 73}, 064013 (2006);
V.~Bozza and G.~Veneziano,
JCAP {\bf 0509}, 007 (2005);
V.~Bozza,
JCAP {\bf 0602}, 009 (2006);
M.~Giovannini,
Phys.\ Rev.\ D {\bf 73}, 083505 (2006).


\bibitem{CNZ05}
P.~Creminelli, A.~Nicolis and M.~Zaldarriaga,
Phys.\ Rev.\ D {\bf 71}, 063505 (2005).

\bibitem{Tolley02}
A.~J.~Tolley and N.~Turok,
Phys.\ Rev.\ D {\bf 66}, 106005 (2002).

\bibitem{Tolley03}
A.~J.~Tolley, N.~Turok and P.~J.~Steinhardt,
Phys.\ Rev.\ D {\bf 69}, 106005 (2004).

\bibitem{Luca}
L.~Amendola,
Phys.\ Rev.\ D {\bf 60}, 043501 (1999);
Phys.\ Rev.\ D {\bf 62}, 043511 (2000);
B.~Gumjudpai, T.~Naskar, M.~Sami and S.~Tsujikawa,
JCAP {\bf 0506}, 007 (2005);
L.~Amendola, M.~Quartin, S.~Tsujikawa and I.~Waga,
Phys.\ Rev.\ D {\bf 74}, 023525 (2006).

\bibitem{ART}
I.~Antoniadis, J.~Rizos and K.~Tamvakis,
Nucl.\ Phys.\ B {\bf 415}, 497 (1994).

\bibitem{PT04}
F.~Piazza and S.~Tsujikawa,
JCAP {\bf 0407}, 004 (2004).

\bibitem{YMO}
H.~Yajima, K.~i.~Maeda and H.~Ohkubo,
Phys.\ Rev.\ D {\bf 62}, 024020 (2000);
A.~Toporensky and S.~Tsujikawa,
Phys.\ Rev.\ D {\bf 65}, 123509 (2002).

\bibitem{Kanti}
P.~Kanti, J.~Rizos and K.~Tamvakis,
Phys.\ Rev.\ D {\bf 59}, 083512 (1999).

\bibitem{Eli}
E.~Elizalde, S.~Nojiri and S.~D.~Odintsov,
Phys.\ Rev.\ D {\bf 70}, 043539 (2004).

\bibitem{NO}
S.~Nojiri and S.~D.~Odintsov,
Phys.\ Lett.\ B {\bf 595}, 1 (2004);
Phys.\ Rev.\ D {\bf 70}, 103522 (2004).

\bibitem{Sri}
S.~K.~Srivastava,
arXiv:hep-th/0411221.

\bibitem{Tre}
P.~Tretyakov, A.~Toporensky, Y.~Shtanov and V.~Sahni,
Class.\ Quant.\ Grav.\  {\bf 23}, 3259 (2006).

\bibitem{SST06}
M.~Sami, P.~Singh and S.~Tsujikawa,
arXiv:gr-qc/0605113, Physical Review D to appear.

\bibitem{curvaton}
D.~H.~Lyth and D.~Wands,
Phys.\ Lett.\ B {\bf 524}, 5 (2002);
K.~Enqvist and M.~S.~Sloth,
Nucl.\ Phys.\ B {\bf 626}, 395 (2002);
T.~Moroi and T.~Takahashi,
Phys.\ Lett.\ B {\bf 522}, 215 (2001)
[Erratum-ibid.\ B {\bf 539}, 303 (2002)].



\end{thebibliography}
\end{document}